\documentclass[12pt,A4]{article}
\usepackage{amsmath}

\usepackage{graphics}
\usepackage[dvips]{graphicx}
\usepackage{authblk}

\begin{document}

\title{A stochastic model of the influence of buffer gas collisions on Mollow spectra}

\author[1,2]{K.~Doan~Quoc}
\author[1,3]{T.~Bui~Dinh}
\author[1]{V.~Cao~Long}
\author[1]{W.~Leo\'nski}
\affil[1]{Quantum Optics and Engineering Division, Institute of Physics,\authorcr
 University of Zielona G\'ora Prof. Z. Szafrana 4a, 65-516 Zielona G\'ora, Poland}
\affil[2]{Quang Tri Teacher Training College, Km3, Highway No. 9, Dong Ha,\authorcr Quang Tri, Vietnam}
\affil[3]{Vinh University, 182 Le Duan Street, Vinh City, Vietnam}

\renewcommand\Authands{ and }

\date{}
\maketitle
\abstract{
In this paper, using the idea introduced in \cite{01} and developed in \cite{02} we consider the influence of collisional fluctuations on the Mollow spectra of resonance fluorescence (RF). The fluctuations are taken into account by a simple shift of the constant detuning, involved in a set of optical Bloch equations by collision frequency noise which is modelled by a two-step random telegraph signal (RTS). We consider in detail the Mollow spectra for RF in the case of an arbitrary detuning of the laser frequency, where the emitter is a member of a statistical ensemble in thermodynamic equilibrium with the buffer gas at temperature $T$ which is treated as a colored environment, and velocity $v$ is distributed with the Maxwell-Boltzmann density.
} 
\newpage
\section{Introduction}
\label{intro}
Research on the fluorescence spectra, which is driven by the application of an intense resonant laser beam, plays a fundamental role in the efforts to achieve a deep understanding of processes in which radiation interacts with matter. The phenomenon of RF of laser-driven atomic multilevel systems and both its spectral as well as statistical, time-resolved properties have been intensively studied for a long time. The RF of atoms and individual atoms in particular, has been the subject of quantum optical measurements for many years \cite{03,04}. A well-known example is the excitation spectrum of a two-level atom interacting resonantly with a strong electromagnetic field, which consists of a central peak at the excitation frequency and two symmetrically placed sidebands, shifted from the central peak position by the Rabi frequency. This well-known spectrum has been predicted theoretically by Mollow \cite{05} and confirmed experimentally by Shuda {\it et al.} \cite{06}, where the correct ratio of the height of the central peak to the height of the side component (3:1) has been determined. This spectrum is related to the second-order correlation function of the fields. Recently, precise measurements of the third-order correlation function of the fields (the intensity-field correlation) \cite{07} and the fourth-order correlation function (the correlation of two intensities) \cite{08} have been performed. In this paper, we study only the Mollow spectrum.
 
	The RF of atoms is a typical problem of quantum optics, where we have the resonance interaction of several lasers with an atomic (molecular) system, the laser lights are fluctuating in amplitude and phase and collisions between elements of the system are frequently taken into account. Because of the very complicated (in many cases even obscure!) microscopic nature of all relevant relaxation mechanisms, they are usually modeled by classical time-dependent random process. Then dynamical equations involved in the problem become stochastic differential equations. Generally they can not be solved in finite terms, especially when the coherence time of the noises (laser light or collisions) is comparable to the intrinsic time scale of the atomic system.  One of the most useful stochastic models has been proposed by W\'odkiewicz and his coworkers \cite{09,10,11,12} which is based on the so-called pregaussian process. It is composed of a finite number of independent identical two-step random telegraph signals (RTS). The strength of the pregaussian formalism derives from the exact solubility of wide classes of the stochastic equations. Moreover the pregaussian noises converge very fast to the colored (chaotic) noise. Even in the case of one (RTS) one can obtain several interesting results \cite{13}.
 
 	The usefulness of RTS is not limited only to quantum optics. As it has been emphasized in \cite{14}, RTS is very useful model for environmental fluctuations which exist in several nanodevices. The authors of the paper \cite{14} considered the dynamics of quantum correlations in colored environments by developing a formalism, in which RTS also is used as the basic building element to describe so-called colored noise being a collection of RTS with different switching rates, so in contrary to the case of pregaussian formalism, the RTS components are not identical. For obtaining the colored spectrum, the single RTS frequency spectral density involved in the problem should be weighted over the switching rates with a proper distribution.
  
	Here we have the same situation if we treat the buffer gas as an environment in respect to the considered atomic system. Long time ago, W\'odkiewicz in his talk \cite{01} emphasized that for the case of collisional fluctuations, the relevant stochastic differential equations can be obtained from the coherent case by a simple shift of the constant detuning by an amount $x(t)$ which is collision frequency noise. Let us see in detail how this procedure works when we have a two-level atom (emitter) with the levels separated by the frequency $\omega_0$, exposed to a resonant radiation field and embedded in a foreign buffer gas (pertubers) \cite{02}. If $\vec k$ is the wave vector of the emitted photon and $\vec v$ is the instantaneous velocity of the emitter, then by Doppler effect the instantaneous frequency of the two-level atom is $\omega_0+x(t)$, where $x(t)=\vec k \vec v =kv\cos\alpha$, $\alpha$ is the angle between $\vec k$ and $\vec v$. One can assume that due to collisions, only the direction of $\vec v$ changes at random. Each time the emitter is assumed to undergo collisions with the pertubers, the angle $\alpha$ suffers an abrupt change of $\pi$ or $-\pi$. In between the collisions the velocity of the emitter remains constant and as a result we can write the $\cos\alpha=(-1)^{n(t)}$, and $n(t)$ is the number of times the noise $x(t)$ changes its state due to collisions. Thus $x(t)$ is a random variable described by a two-step RTS, with $a = kv$ and the switching rate related to the velocity $1/\tau_a=\pi n_pr_0^2v$ in which $n_p$ is the number of pertubers per unit volume and $r_0$ is an effective scattering radius. In thermodynamic equilibrium with the buffer gas at temperature $T$ treated as a colored environment, and velocity $v$ is distributed with the Maxwell-Boltzman density, the spectrum of emitted light must be weighted with this distribution.

	In the next section we consider the scattered light spectrum of a nondegenerate two-level atom resonantly driven by strong monochromatic laser light in the presence of energy-shifting collisions modeled by the RTS.
\section{Spectrum of the Resonance Fluorescence}
\label{sec:1}
Starting from a set of optical Bloch equations derived from nonrelativistic quantum electrodynamics in the rotating wave approximation (RWA), and shifting the atom energy by the RTS $x(t)$ according to the idea described in Introduction, we can perform an averaging procedure over this RTS and obtain the following formula for the spectrum of RF in the presence of the collisions with the molecules of the buffer gas (since the volume of this communication is limited,  we shall concentrate here only on physical aspects of the problem, whereas the mathematical details of this procedure will be presented elsewhere \cite{15}):
\begin{equation}\label{1}
\begin{split}
S_{RTS}(v,\omega)&=2\text{Re}\Bigg\{\frac{(z+2\gamma)(z+\gamma+\Gamma_{11}+i\Delta)+\frac{\Omega_0^2}{2}}{\big(z+\gamma+\frac{1}{2}(\Gamma_{11}+\Gamma_{33})+\Gamma_{13}+i\Delta\big)}\\
&\times\frac{1}{\Big[(z+2\gamma)(z+\gamma+\frac{1}{2}(\Gamma_{11}+\Gamma_{33})-\Gamma_{13}-i\Delta)+\Omega_0^2\Big]+N}\Bigg\}\Bigg|_{z=i\omega},
\end{split}
\end{equation}
where
\begin{equation}\label{2}
\begin{split}
&\Gamma_{11}=\frac{a^2}{P}\bigg(z+\frac{1}{\tau_a}\bigg)\bigg[\Big(z+\gamma+\frac{1}{\tau_a}-i\Delta\Big)\Big(z+2\gamma+\frac{1}{\tau_a}\Big)+\frac{\Omega_0^2}{2}\bigg],\\
&\Gamma_{33}=\frac{a^2}{P}\bigg(z+\frac{1}{\tau_a}\bigg)\bigg[\Big(z+\gamma+\frac{1}{\tau_a}+i\Delta\Big)\Big(z+2\gamma+\frac{1}{\tau_a}\Big)+\frac{\Omega_0^2}{2}\bigg],\\
&\Gamma_{13}=\Gamma_{31}=-\frac{a^2}{2P}\bigg(z+\frac{1}{\tau_a}\bigg)\Omega_0^2,\\
\end{split}
\end{equation}
\begin{equation}\label{3}
P=\bigg(z+\frac{1}{\tau_a}\bigg)\bigg\{\Big(z+\gamma+\frac{1}{\tau_a}+i\Delta\Big)\Big[\Big(z+\gamma+\frac{1}{\tau_a}-i\Delta\Big)\Big(z+2\gamma+\frac{1}{\tau_a}\Big)+\Omega_0^2\Big]-i\Delta\Omega_0^2\bigg\},
\end{equation}
\begin{equation}\label{4}
N=-\Big\{(z+2\gamma)\Big[\frac{1}{4}(\Gamma_{11}-\Gamma_{33})^2-i\Delta(\Gamma_{33}+2\Gamma_{13}-\Gamma_{11})\Big]+i\Delta\Omega_0^2\Big\}.
\end{equation}
Here $\Omega_0$ is the Rabi frequency, $\Delta=\omega_0-\omega_L$ ($\omega_0$ is the atomic frequency, $\omega_L$ is the frequency of the single-mode driving source of the light), $z$ is the argument of Laplace transform of the second-order correlation function for the fields (in the formula (\ref{1}) we must put $z=i\omega$), $1/\tau_a=\pi bv$ is the RTS rate in which $b=n_pr_0^2$, $n_p$ is the number of pertubers per unit volume and $r_0$ is an effective scattering radius, $v$ denote the speed of a molecule, $a=kv$ in which $k$ is the magnitude of wave vector of the emitted photon, $\gamma=A/2$ where $A$ is the Einstein coefficient.

	If atoms of the buffer gas have different velocities, then spectrum should be averaged with respect to the Maxwell-Boltzmann distribution:
\begin{equation}
S(\omega)=\int_0^\infty \text{d}vf(v)S_{RTS}(v,\omega),\label{5}
\end{equation}
where
\begin{equation}
f(v)=4\pi\Bigg(\frac{c}{2\pi}\Bigg)^{3/2}v^2e^{-\frac{cv^2}{2}},\label{6}
\end{equation}
in which $c=m/k_BT$, $m$ is the mass of a molecule of the buffer gas, $k_B$ is the Boltzmann constant, $T$ is the temperature.

\begin{figure}[!h]
    \begin{center}
\includegraphics[height=3.9cm, width=6cm]{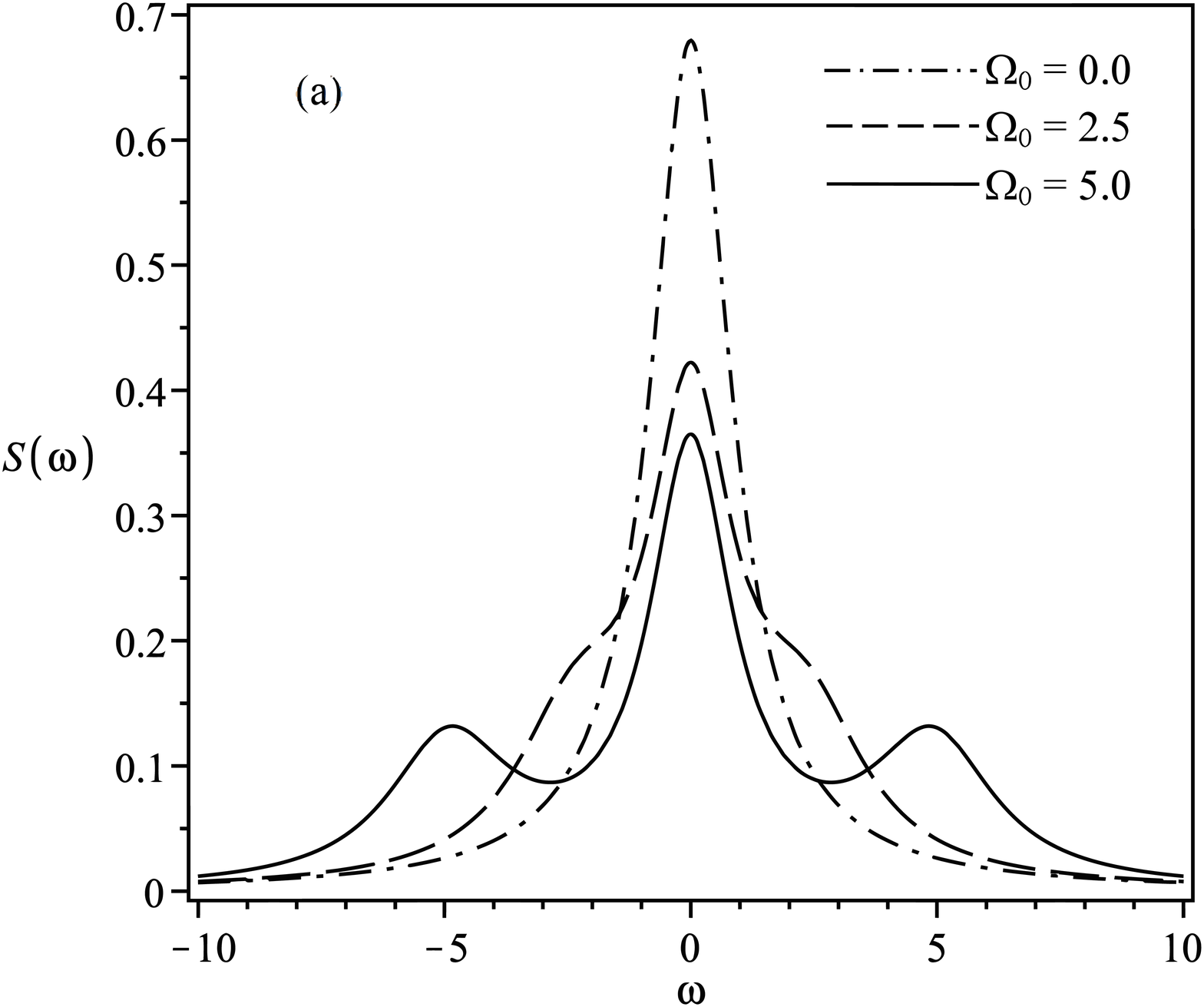}\quad
\includegraphics[height=3.9cm, width=6cm]{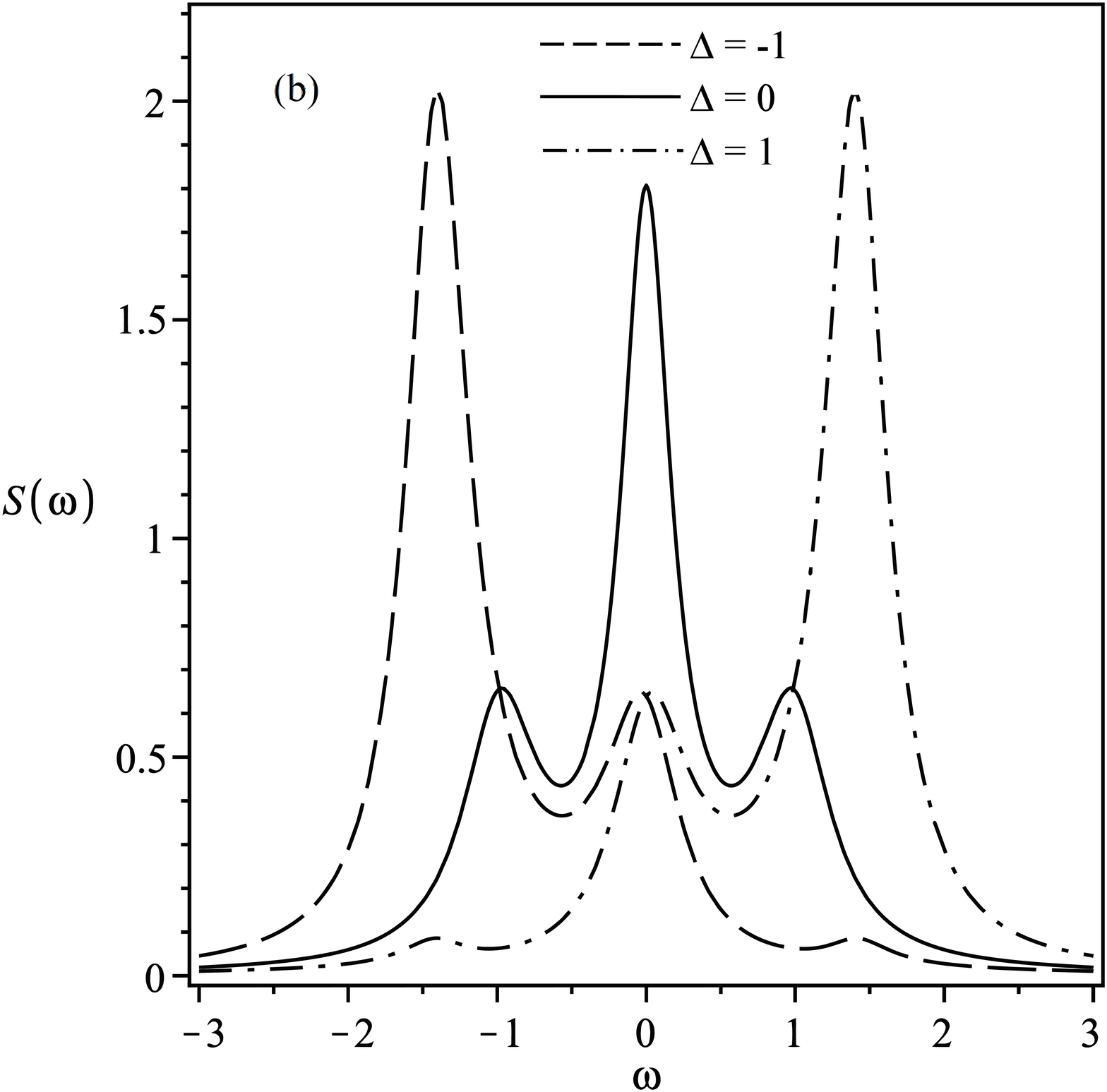}
    \end{center}
\caption{The dependence of the spectrum of resonance fluorescence on different values of (a) Rabi frequency in the exact resonance $(\Delta=0)$ or (b) $\Delta$ with $\gamma=0.2$. For both (a) and (b) $k=0.01$, $b=0.9$ and $c=1.6\times 10^{-4}$.}
\label{f1}       
\end{figure}

\begin{figure}[!h]
    \begin{center}
\includegraphics[height=3.9cm, width=6cm]{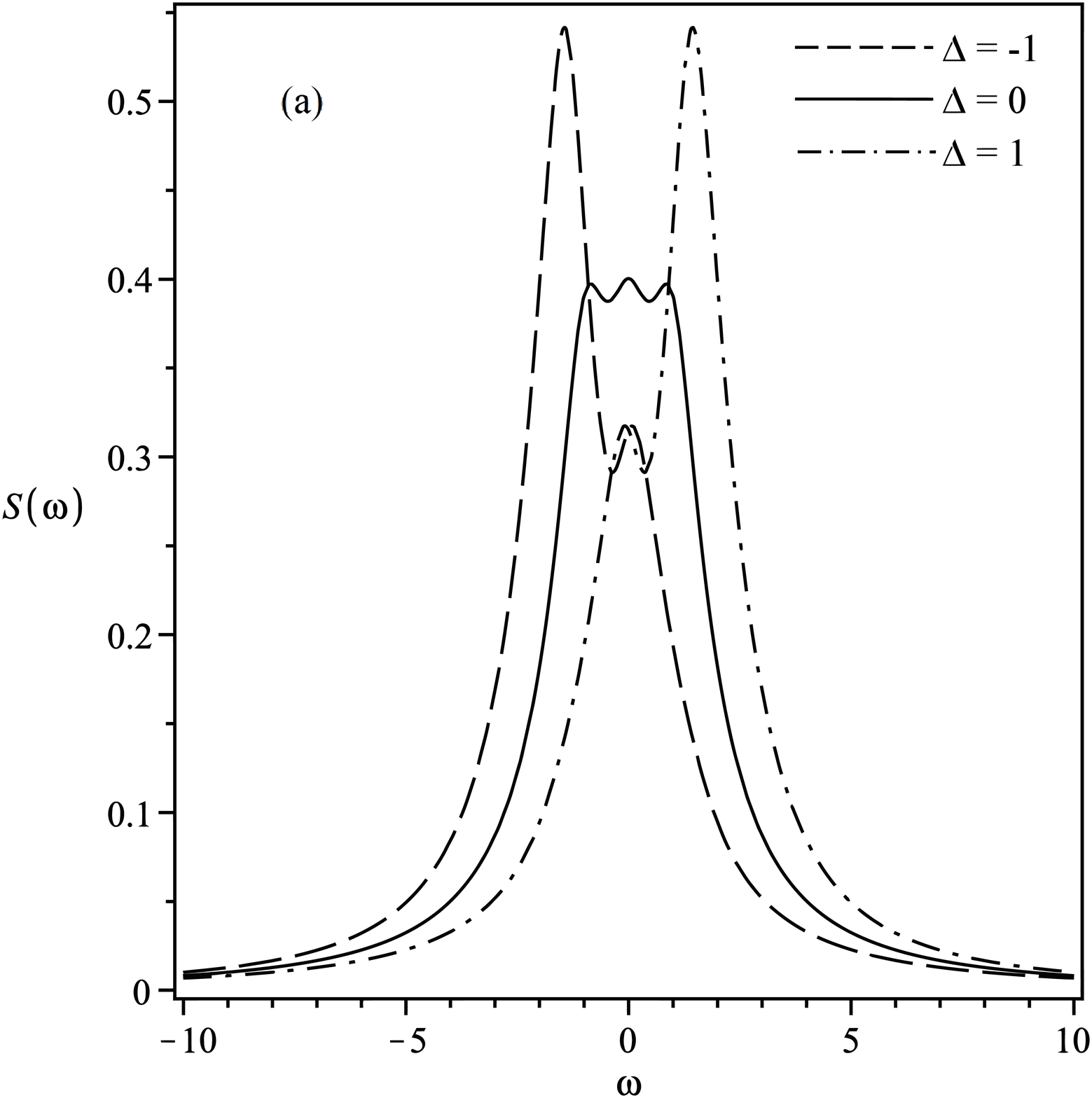}\quad
\includegraphics[height=3.9cm, width=6cm]{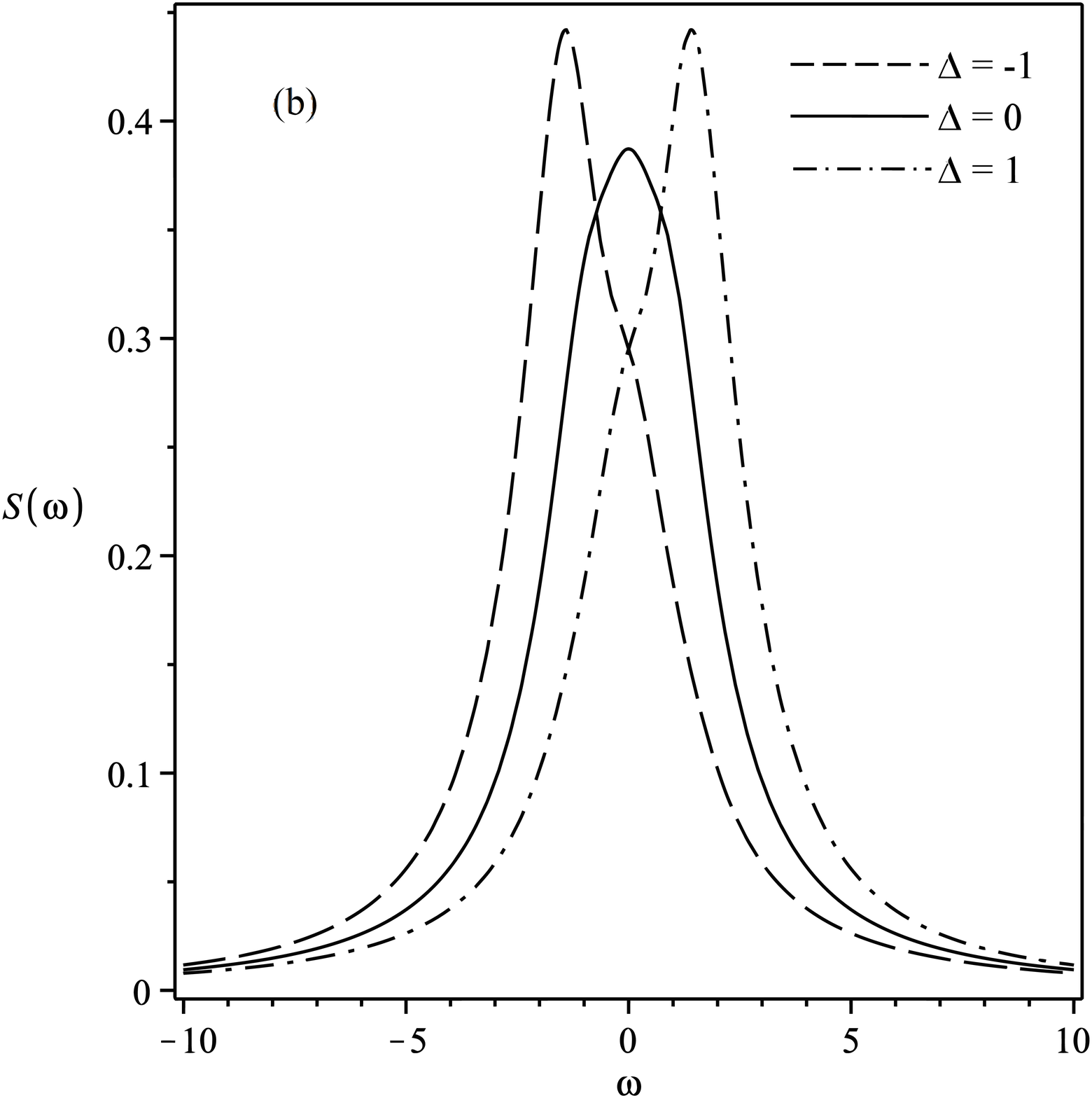}
    \end{center}
\caption{The dependence of the spectrum of resonance fluorescence on different values of $\Delta$ for $k=0.2$, $b=0.9$ and $c=1.6\times 10^{-4}$ in which (a) $\gamma=0.2$ and (b) $\gamma=0.4$.}
\label{f2}       
\end{figure}

	We calculate the integral (\ref{5}) numerically for different values of the parameters involved in the problem. The results are presented in figures \ref{f1} and \ref{f2}. Figure \ref{f1}a uses $\gamma$ unit, whereas figures \ref{f1}b and \ref{f2} use $\Omega_0$ unit. The quantities $k$ and $b$ are in unit $r_0^{-1}$ and $c$ is in unit $1/v_0^2$, $v_0=\gamma r_0$. The collisional fluctuations always occur in all figures which are presented here.

When the detuning is equal to zero $(\Delta=0)$, we obtain the well-known Mollow spectrum (figure \ref{f1}a), which consists of a symmetric triplet, whose side components are separated from the central peak by a distance equal to the generalized Rabi frequency, but the amplitude is not equal to one third of that of the central peak because chaotic component is present $(k\neq 0)$. When collisions are taken into account, the centre of gravity of the triplet shifts towards the side bands, in contrast to the prediction of \cite{16,17}, where the centre of gravity of the multiplet shifts towards the frequency of the resonance transition. For larger collisional fluctuations, the maximum values of the peaks are lower. As long as the detuning is present $(\Delta\neq 0)$, the spectrum of RF is asymmetric. However, we consider the dependence of the spectrum on the detuning for the fixed values of other parameters. As in our recent paper, the theory of RF is commonly presented in the literature within the framework of the RWA. This is justified in the case of a driving field frequency that is nearly resonant with the atomic transition frequency, assuming that the driving frequency is significantly larger than the Rabi frequency. It is commonly accepted that if a detuning is present, the theory must be modified, then we can not use the RWA. This certainly leads to asymmetric spectra \cite{18}. Here, we still apply the RWA and consider the dependence of the spectrum on the detuning while keeping other parameters fixed.

For $\gamma=0.2$, the spectrum is presented in the figure \ref{f1}b. We obtain a three-peak asymmetric spectrum. When $\Delta=-1$, the central peak decreases, whereas the wing components change in the opposite way (the left peak increases, whereas the right decreases). When $\Delta=0$, the spectrum is symmetry in which central peak is highest. When $\Delta=1$, the central peak decreases, so does the left wing, while the right wing increases.

	When collisional fluctuations are large, the intensities of all peaks are smaller than for the small collisional fluctuations and the three-peak structure of the spectrum is progressively destroyed (see figure \ref{f2} for $k=0.2$).

	Moreover, when the spontaneous emission rate $\gamma$ increases, the three-peak structure of the spectrum is also progressively destroyed (see figure \ref{f2}).
\section{Summary}
\label{sec:2}
We have presented a stochastic approach to the incoherence properties induced by collisions. Our treatment was based on modeling collisions by random telegraph noises. It has been shown that when the detuning is equal to zero, we obtain the well-known Mollow triplet. When the collisional fluctuations increase, the centre of gravity of the triplet shifts towards the side bands, in contrast to the prediction in \cite{16,17}. Furthermore, the triplet remains symmetric, so the asymmetry of the fluorescence spectra found experimentally in \cite{19} is dictated by the fundamentally multilevel nature of the atomic system, and not by the collisions. If a detuning is present, the theory must be modified without the use of RWA. This leads to asymmetric spectra \cite{18} with emerging additional harmonics. 
%
%
%
%

\end{document}